\documentstyle[11pt]{article}
\normalbaselines
\begin{document}
\bibliographystyle{unsrt}

\vbox{\vspace{38mm}}
\begin{center}
{\LARGE \bf A NEW SPINNING MEMBRANE LAGRANGIAN submitted to the\\[2mm]
Journal of Group Theory in Physics}\\[5 mm]

Carlos Castro \\{I.A.E.C 1407 Alegria\\
Austin, Texas 1407 USA }\\[3mm]
(January 1990; Submitted  March 1993)\\[5mm]
\end{center}

\begin{abstract}
\bigskip
   A new local world volume supersymmetric Lagrangian for the bosonic
membrane is presented. The starting Lagrangian is the one constructed by
Dolan and Tchrakian with vanishing cosmological constant, with quadratic and
quartic derivative terms. Our Lagrangian differs from the one constructed
by Lindstrom and Rocek in the fact that it is polynomial in the fields
facilitating the quantization process. It is argued, rigorously, that if one
wishes to construct polynomial actions without a curvature scalar
term and, where supersymmetry is linearly realized in the space of physical
fields, after the elimination of auxiliary fields, one must relinquish $S$
supersymmetry, altogether, and concentrate solely on the  $Q$ supersymmetry
associated with the superconformal algebra in three dimensions. A full
$''Q+S''$ supersymmetry cannot be implemented in a linearly realized way
satisfying all of the above-mentioned requirements, unless a non-polynomial
action is chosen.


\end{abstract}

\section {Introduction}

      In the past years there has been considerable
progress in the theory of two dimensional extended
objects; i.e. membranes. However, a satisfactory spinning
membrane Lagrangian has not been constructed yet, as far
as we know. Satisfactory in the sense that a suitable
action must be one which is polynomial in the fields,
without $ R$ terms (curvature scalar) which interfere with the algebraic
elimination of the three-metric, and also where
supersymmetry is linearly realized in the space of
physical fields. Howe, Duff and others \cite{howe 1}
\cite{duff 2}
sometime ago mentioned the fact that it is allegedly impossible to
supersymmetrize
Dirac-Nambu-Goto type of actions (DNG)
-those proportional to the world-sheet and
world-volume spanned by the string (membrane) in their
motion through an embedding space-time. The efforts to
supersymmetrize this action have generally been based upon
the use of the standard, classically-equivalent, bosonic
action which included a cosmological constant. The
supposed obstruction is related to the fact that in order
to supersymmetrize this constant one had to include an
Einstein-Hilbert term spoiling the process altogether.

       Bergshoeff et al \cite{berg 3}  went evenfurther and
presented us with the "no-go" theorem for the spinning membrane. Their finding
was based in the study of a family of actions, in addition to the one comprised
of the cosmological constant, which were equivalent, at the
classical level,to the DNG action. However, this "no-go"
theorem was flawed because these authors relied on the
tensor calculus for Poincare D=3 N=1 SG developed by
Uematsu. \cite{uematsu 4}
Unfortunately, the above tensor calculus does not even yield a linearly
realized
supersymmetry for the kinetic matter multiplet to start with!. A constraint,
 $\bar{\chi}\chi$ =0  ,  appears after the elimination
of the S auxiliary field where $\chi$ is the three dimensional Majorana spinor.
 By the spinning membrane one means that one has supersymmetry on the
world-volume
whereas by the supermembrane one means that one has supersymmetry on the
embedding spacetime
background.  Lindstrom and Rocek \cite{lindstrom 5}
were the first ones to construct a Weyl invariant spinning membrane action.
However, such action was highly non-polynomial
complicating the quantization process evenfurther than the
one for the supermembrane.

     The suitable action to supersymmetrize is the one of
Dolan and Tchrakian \cite{dolan 6}  without a cosmological constant and with
quadratic and
quartic-derivative terms. A class of conformally-invariant $\sigma$- model
actions was shown to be equivalent, at the classical level, to the
DNG action for a $p+1$ extended object ( $p+1$=even) embedded
in a target spacetime of dimension $ d\geq {p+1}$. When
p+1=odd, our case, an equivalent action was also
constructed, however, conformal invariance was lost in
this case.
The crux of this work lies on the necessity to embed the Dolan-Tchrakian
action in a Weyl-invariant one through the introduction of extra fields. These
are the gauge fields of dilations, $b_\mu$, and the scalar coupling of
dimension
$(length)^{+3}$, $A_0$, which appears in front of the quartic derivative terms.
The latter must appear with a suitable coupling constant in order to render
the action dimensionless. Upon embedding the Dolan-Tchrakian action in a
Weyl invariant one this coupling constant becomes a scalar. A similar
procedure occurs in the Brans-Dicke formulation of gravity.

   Having gone through the embedding procedure the natural question to ask
is how do we eliminate these new fields, $b_\mu,A_o$, in order to recover
the original action?    One will recover back the Dolan-Tcrakian action by
fixing the dilational invariance :$A_0=constant$ and enforcing $b_\mu=0$.
This is achieved, simultaneously,  if one imposes
the natural constraint on $A_0$: $D^{Weyl}_\mu A_0=\partial_\mu A_o +3b_\mu
A_o =0$. Such constraint can
vbe derived from first principles: from an action. It follows automatically
that if the equations of motion of the Weyl-covariantized Dolan-Tchrakian
 action are, indeed,the Weyl-covariant extension of the Dolan-Tchrakian
 equations of motion, then $D_\mu A_o =0$ follows immediately.
 The reason why we cannot fix the conformal boost gauge invariance by
setting :$b_\mu =0$, is because our
final spinning membrane action  is not invariant under conformal boosts
(the $b_\mu$ field does not decouple) nor under $S$ supersymmetry. Therefore,
 on-shell dilational gauge invariance of the Weyl
covariantized Dolan Tchrakian (WCDT) action allows us to recover the
original DT action. Notice that we have not imposed  any constraints,
 whatsoever, on our physical fields!. (See Appendix for further details).

    Once the embedding process is performed one supersymmetrizes the WCDT
action by incorporating $A_o$ into a superconformal coupling-function
multiplet $(A_o,\chi_o,F_o)$. The field $b_\mu$ is part of the
superconformal gauge multiplet involving $(e^m_\mu,\psi_\mu,b_\mu)$, and
the physical fields of the membrane form part of the superconformal matter
multiplet, $(A,\chi,F)$. The $A^i$ fields are identified with the
membrane's embedding coordinates, $X^i$.  Finally, if we wish to eliminate any
curvature
scalar terms one must take suitable combinations of products involving
these three superconformal multiplets and, in doing so, one is going to
break, explicitly, the $S$ supersymmetry of the three-dim superconformal
algebra as well as the conformal boost symmetry, the $K$ symmetry, which
signals the presence of the $b_\mu$ field in our final spinning membrane
action.

    The final action is Lorentz, dilational, $Q$ supersymmetric and
translational invariant. There is nothing wrong with these fact since the
subalgebra of the full three-dim superconformal algebra comprised of the
Lorentz generator, dilations, $Q$ supersymmetry and translations, $P_\mu$,
does indeed close !! In conventional Poincare supergravity one has an
invariance under a particular linear combination of $Q$ and $S$
supersymmetry : the so-called $Q+S$ rule as well as $K$ symmetry which
enforces the decoupling of the $b_\mu$ field from the supersymmetric
action. Here we have a different picture, we have full $Q$ supersymmetry,
instead of a particular combination of $Q$ and $S$, and no conformal boost
invariance. This is the main peculiarity of our spinning membrane action
and the most important result from the group theoretical point of view.

  The outline of the paper goes as follows:

       In section II  we discuss the work of Dolan and
Tchrakian (DT) and show the equivalence to DNG type of actions. In section  III
we present the problems associated with the Poincare tensor calculus and point
out why the "no-go" theorem was inappropriate. In the final
section, IV, we give a detail argument showing that in
order to satisfy all of the stringent requirements
discussed earlier, we must relinquish $ S$-supersymmetry and
concentrate, solely, on the $ Q$-supersymmetry associated with
the superconformal algebra in three dimensions. $"Q+S$"
supersymmetry can only be implemented in the class of
non-polynomial actions (if we insist in meeting all of our
requirements) as it was shown by Lindstrom and Rocek . Finally, the fully
$Q$-invariant action is presented in $4.1$; In subsection $4.2$ we discuss
in  full detail how to retrieve the original Dolan-Tchrakian action, after
eliminating the auxiliary fields and setting the fermions to zero, and
why, then, we have a $Q$ spinning membrane. An
appendix is included where we derive from first principles the embedding
condition :$D_\mu A_o =0$, which enables us to set $b_\mu =0,$ while fixing
$A_o$ to a constant recovering in the process the original DT action.

Our coventions are: Greek indices, $\mu,\nu...$ stand for
three-dimensional world volume ones: $ 0,1,2$ ; Latin indices, $i,j,k...$ for
 spacetime ones.

\section {The Dolan-Tchrakian Action}

    The new Lagrangian for the bosonic m-extendon (m-brane) with vanishing
cosmological constant constructed by Dolan and Tchrakian  for $m=odd.~~m+1
=2n$, is :
$$L_{2n} =\sqrt {-g} g^{\mu_1 \nu_1 }........g^{\mu_n \nu_n }
\partial_{[\mu_1} X^{i_1}......\partial_{\mu_n ]}X^{i_n}
\partial_{[\nu_1} X^{j_1}......\partial_{\nu_n ]}X^{j_n} \eta_{i_1 j_1}....
\eta_{i_n j_n}. \eqno(2-1) $$

$\eta_{ij}$ is the spacetime metric and $g^{\mu \nu}$ is the metric for the
$2n$
dimensional hypersurface spanned by the m-extendon and the antisymmetrization
of indices is explicitly shown. Upon elimination of the $2n\times 2n$ matrix :

$$ A^\mu_\nu =g^{\mu \rho} \partial_\rho X^i (\sigma) \partial_\nu X^j (\sigma)
\eta_{ij}. \eqno (2-2)$$

 from the $n^{th}$ order polynomial in $A$ :
$$ A^n -b_{2n -1} A^{n-1} +b_{2n -2} A^{n-2} -.........(-1)^{n+1} b_{n+1} A +
1/2~(-1)^n b_n I_{2n \times 2n} =0. \eqno (2-3)$$
where the scalar coefifcients $b^i$ are the same as the first $n+1$
coefficients in the expansion :
$$det (A-\lambda I) =\lambda ^{2n} -b_{2n-1}\lambda ^{2n-1} +b_{2n-2}
\lambda^{2n-2} .......-b_1 \lambda +det A. \eqno(2-4)$$
and substituting the solution of (2-3) back into (2-1) one finds :

$$L_{2n} =\sqrt {-det (\partial_\mu X^i \partial_\nu X^j \eta_{ij} )} [(n!)^2
b_n /\sqrt A]. \eqno(2-5)$$

    The crucial observation made by Dolan and Tchrakian was that the last
factor
in (2-5) takes discrete values for all $n$. Therefore, the equivalence to the
Nambu-Goto action. Notice that for every $n$, $L_{2n}$ is conformally invariant
under scalings and despite $L_{2n}$ being of order $2n$ in derivatives it is
only
quadratic in time derivatives because of the antisymmetry in the wedge type of
product in (2-1). Therefore, attempts to quantization might not be hopeless.

\smallskip

    When $m=even; m+1 =2n+1$ a Lagrangian with zero cosmological constant can
also be constructed; however, conformal invariance is lost. For our case, the
membrane. $m=2$ the Lagrangian is :
$$L_{mem} =L_4 +aL_2.\eqno(2-6)$$
$$L_4 =\sqrt {-g} g^{\mu\nu} g^{\rho \tau}
\partial_{[\mu} X^i \partial_{\rho ]}X^k \partial_{[\nu } X^j
 \partial_{\tau ]}X^l \eta_{ij} \eta_{kl}. \eqno(2-7)$$
and

$$L_2 =\sqrt {-g} g^{\mu\nu}\partial_\mu X^i \partial_\nu X^j \eta_{ij}.\eqno
(2-8)$$

where $\mu,\nu =0,1,2; a=constant;i,j...=1,2,....d$. The above Lagrangian upon
elimination of the world volume metric is :
$$12 \sqrt a~\sqrt {-det G} ~or~-4\sqrt a~\sqrt {-det G}.\eqno (2-9)$$
with $G_{\mu\nu} =\partial_\mu X^i \partial_\nu X^j \eta_{ij}.$

\section {A nonlinearly realized supersymmetric membrane}
    We shall begin by writing down the supersymmetric term (modulo total
derivatives) for the Poincare kinetic scalar multiplet in D=3 N=1 SG given by
Uematsu :

$$L={1\over2}[\Sigma_P\otimes T(\Sigma_P)]_{inv}
-{1\over4}[T(\Sigma_P\otimes \Sigma_P)]_{inv} =$$
$$e[-1/2~g^{\mu \nu} \partial_\mu A\partial_\nu A -1/2~\bar {\chi}\gamma^\mu
D_\mu \chi +1/2~F'^2 +1/2~\bar {\psi_\nu} \gamma^\mu \gamma^\nu
\chi\partial_\mu A +$$
$$1/16~\bar {\chi} \chi \bar {\psi_\nu}\gamma^\mu \gamma^\nu \psi_\mu
+1/8~S\bar {\chi} \chi]. \eqno(3-1)$$

The kinetic multiplet -which is not uniquely defined since it is defined up to
the addition of a fixed scalar multiplet which starts with SA as its first
component- is given by:
$$ \tilde A = F'.~~~{\tilde \chi} =\gamma^\mu D^P_\mu \chi -1/4~S\chi. \eqno
(3-2a,b)$$

 $$\tilde F' =e^{-1}\partial_\mu (eg^{\mu \nu}D^P_\nu A ) -1/2~\bar
{\psi^\nu}\gamma^\mu \psi_\mu D^P_\nu A -1/2~\bar {\psi^\mu} D^P_\mu \chi +$$
$$i/4~e^{-1}\epsilon^{\mu\nu\rho} \bar {\chi} \gamma_\rho \psi_{\mu \nu}
-F'S+1/8~S\bar {\psi_\mu }\gamma^\mu \chi. \eqno(3-2c)$$
\smallskip
     Notice that (3-1) is ,by itself, unsatisfactory because upon elimination
of $S$ via its equations of motion one ends up with an unnatural constraint
 among
the physical fields of the theory ,$\bar {\chi} \chi =0$. Similar results are
obtained when we supersymmetrize the quartic terms. \smallskip
     The appearace of such constraint after the elimination of $S$ traces all
the way back to the tensor calculus and transformation rules for Poincare D=3
 N=1 SG given by Uematsu. Such rules are essentially identical to the
two-dim case except for a minor modification in the transformation law of the
matter auxiliary field, $F'$. This, in turn, forces an extra term, linear in
$S$, in the ${\tilde \chi};\tilde F'$ components of the kinetic multiplet.
Armed
with these minor changes, one obtains (3-1). These modifications "propagate"
to the quartic terms also and unwanted linear couplings among $S$ and the
matter
fields appear forcing unnatural constraints after the elimination of $S$. In
order to remedy this one could add the pure Supergravity cation with a
corresponding $S^2$ term. However, this is precisely what we wanted to avoid :
the presence of $R$ terms in our action !.\smallskip

    There are ways to circumvent this problem. One way was achieved by Linstrom
and Rocek who started from a non-polynomial but conformally invariant
action
 : $$I\sim \int d^3\sigma~\sqrt {-g} (g^{mn}\partial_m X^\mu \partial_n X^\nu
\eta_{\mu\nu} )^{3/2}. \eqno (3-3)$$

where $m,n =0,1,2.~~~\mu ,\nu =1,2.......d.$ Since $S$ is an "alien" concept in
conformal supergravity, it cannot appear in the supersymmetrization process,
whether one uses  Conformal or Poincare supergravity techniques to build
invariant actions. This was explictly verified by Lindstrom and Rocek. The
shortcoming is that the action is highly non-polynomial complicating
evenfurther
the quantization process than the one of the supermembrane. On the other hand,
the Dolan-Tchrakian action is polynomial but as a result of the non-linearly
realized supersymmetry due to the matter fields constraints (upon elimination
of $S$) the quantization programme is going to be hampered considerably.
In the next section we present ways to supersymmetrize the
Dolan-Tchrakian action. The crucial difference is that we shall only implement
$Q$ supersymmetry  instead of both $Q$ and $S$ supersymmetry of the
superconformal algebra in three dimensions.

\section{The Lagrangian}
\subsection{A $Q$ Spinning Membrane}

    In this section we will present an action for the 3-dim-\ Kinetic matter
supermultiplet where supersymmetry is linearly realized and without R terms.
Also we will supersymmetrize the quartic-derivative terms of (1-1). This is
attained by using directly an explicitly superconformally invariant action for
the kinetic terms. The quartic terms do not admit a superconformally invariant
extension as we shall see shortly. The key issue lies in the fact that if we
wish to satisfy the three requirements:

1). A spinning membrane action which is polynomial in the fields.
\smallskip

2) Absence of R terms.
\smallskip

3). Linearly realized supersymmetry in the space of fields after the
elimination of the auxiliary fields, before and after one sets the Fermi fields
to zero.

   One must relinquish S-supersymmetry altogether and concentrate solely on the
Q-supersymmetry associated with the superconformal algebra in D=3. We shall
begin with some definitions of simple-conformal SG in D=3. [ 4 ]:

The scalar and kinetic multiplet of simple conformal SG in D=3 are
respectively:

$$\Sigma_c =(A,\chi,F).$$ $$T_c(\Sigma_c)=(F,\not {D}^c\chi,\Box^cA)
\eqno(4-1)$$

We have the following quantities:

$$ D^c_\mu A=\partial_\mu A -1/2\bar{\psi}\chi -\lambda b_\mu A.
                                   \eqno(4-2)$$

$$ D^c_\mu\chi=(D_\mu-(\lambda+1/2)b_\mu )\chi -1/2 \not D^c A\psi_\mu
-1/2F\psi_\mu -\lambda A\phi_\mu.
\eqno(4-3)$$

$$ \Box^c A=D^c_aD^{ca} A=e^{-1}\partial_\nu(eg^{\mu\nu}D^c_\mu A)+1/2\bar
{\phi_\mu}\gamma^\mu \chi -(\lambda -1)b^\mu D^c_\mu A+$$
$$2\lambda A f^a_\mu e^\mu_a -1/2\bar{\psi^\mu}D^c_\mu \chi
-1/2\bar{\psi^\mu}\gamma^\nu\psi_\nu D^c_\mu A.               \eqno(4-4)$$

$$\omega^{mn}_\mu =-\omega^{mn}_\mu (e)-\kappa^{mn}_\mu(\psi)+e^n_\mu b^m
-e^m_\mu b^n.                                              \eqno(4-5-a)$$

$$\phi_\mu =1/4\gamma^\lambda \gamma^\sigma \gamma_\mu S_{\sigma\lambda}
  =1/2\sigma^{\lambda\sigma}\gamma_\mu S_{\sigma\lambda}.
   \eqno(4-5-b)$$

$$\kappa^{mn}_\mu =1/4(\bar\psi_\mu
\gamma^m\psi^n-\bar\psi_\mu\gamma^n\psi^m +\bar\psi^m\gamma_\mu\psi^n).$$

$$S_{\mu\nu}=(D_\nu +1/2~b_\nu )\psi_\mu  -\mu \leftrightarrow \nu.
\eqno(4-5-d)$$

$$e^{a\mu}f_{a\mu}=-1/8R(e,\omega)-1/4\bar\psi_\mu \sigma^{\mu\nu}\phi_\nu.
\eqno(4-5-e)$$

         The transformation laws under Weyl scalings, $Q$ and $S$ supersymmetry
are
respectively:

$$\delta e^m_\mu =\lambda e^m_\mu ;~~\delta g^{\mu\nu}=2\lambda
g^{\mu\nu};~~\delta A=1/2 \lambda A;~~\delta \chi =\lambda \chi ;\delta
F=3/2\lambda F                                        \eqno(4-6)$$

$$\delta^c_Q e^m_\mu =\bar{\epsilon}\gamma^m\psi_\mu;~~\delta^c_Q
\psi_\mu=2(D_\mu +1/2b_\mu )\epsilon;~~\delta^c_Q b_\mu =\phi_\mu
              \eqno(4-7a)$$
$$\delta^c_Q A=\bar{\epsilon}\chi;~~\delta^c_Q \chi =F\epsilon +\not D^c
A\epsilon.~~ \delta^c_Q F=\bar{\epsilon}\not D^c\chi.
                     \eqno(4-7b)$$

$${\delta^c_S b_\mu ={-1\over 2}\psi_\mu \epsilon_s.}\eqno(4-8a)$$

$${\delta^c_S \omega^{mn}_\mu =- \epsilon_s \sigma^{mn} \psi_\mu.}\eqno(4-8b)$$

$$\delta^c_S e^m_\mu =0;~~\delta^c_S \psi_\mu =-\gamma_\mu \epsilon_s \eqno
(4-8c).$$

$$\delta^c_S A=0;~~\delta^c_S \chi =\lambda A\epsilon_s;~~\delta^c_S
F=(1/2-\lambda)\bar{\chi}\epsilon_s .\eqno(4-8d)$$

    Notice that the kinetic multiplet transforms propely under
$Q$-transformations for any value of the conformal weight, $\lambda$, but
 not under $S$ transformations unless one assigns the canonical weight
 $ \lambda_c=  {1\over 2}$.
Furthermore, if we wish to write down superconformally invariant actions [4]
for a kinetic multiplet, we must have for Lagrangian :
$${L=e[F+{1\over2}\bar{\psi_\mu}\gamma^\mu\chi
+{1\over2}A\bar{\psi_\mu}\sigma^{\mu\nu}\psi_\nu].}\eqno(4-9)$$

and  make sure to have built the kinetic multiplet from a matter multiplet
whose  $ \lambda  = { 1\over 2} $ otherwise we would not
even have Q-invariance  despite the fact that the kinetic multiplet transforms
properly under Q-transformations irrespectively of the value of $\lambda $.
On physical grounds we see that the notion of canonical dimension is
intrinsically tied up with the conformal invariant aspect of the kinetic terms
in the action. We have a conformally invariant kinetic term if,and only if, the
fields have the right (canonical)dimensions to yield terms of dimension three
in
the Lagrangian. We might ask ourselves how did Lindstrom \& Rocek manage to
construct a Weyl invariant spinning membrane when their fields had a
non-canonical dimension? The answer to this question lies on the nonpolynomial
character of their action. Formally one has an infinite series expansion where
each explicitly $Q$ and $S$-breaking term is cancelled by the next term in the
expansion.
   The task now is to see how do we write a suitable action for the kinetic
terms without R terms ( which appear in the definition of the D'Alambertian)
for values of $\lambda$ different than zero. The suitable action is obtained
as follows:

Take the   combination $\Sigma^i_C \otimes T_C(\Sigma^j_C ) + T_C(\Sigma^i_C
)\otimes \Sigma^j_C -
   T_C( \Sigma^i_C \otimes \Sigma^j_C)$ which happens to be the correct one to
dispense of the R terms. The explicit components of the latter multiplet are
(Notice the  $b\mu$ terms):

$${A= \bar{\chi_i} \chi_j.}\eqno(4-10)$$

$${\chi =F_i\chi_j +F_j\chi_i -{1\over2}\bar{\chi_i}\chi_j\gamma^\mu
\psi_\mu -\not \partial A_i\chi_j -\not \partial A_j \chi_i.}\eqno(4-11)$$

$$F= -2g^{\mu\nu}\partial A_i\partial A_j -2\bar{\chi_i}\not D\chi_j
+F_iF_j+{1\over2}\bar{\chi_i}\gamma^\mu \gamma^\nu \psi_\mu \partial_\nu A_j$$
$${+({i\leftrightarrow j}) +\bar{\chi_i}\psi^\mu \partial_\mu
A_j+({i\leftrightarrow j})- {1\over2} \bar{\chi_i}
\chi_j\bar{\psi_\nu}\gamma^\mu  \gamma^\nu\psi_\mu+{1\over4} \bar{\chi_i}\chi_j
\bar{\psi^\mu}\psi_\mu}$$
$$+{1\over2}\bar{\chi^i}\gamma^\mu\psi_\mu F^j+({i\leftrightarrow j})
+{1\over2}\bar{\chi^i}\gamma^\mu\phi_\mu A^j
+({i\leftrightarrow j})$$
$$+\lambda e^{-1}\partial_\nu (eg^{\mu\nu}b_\mu A_iA_j)+\lambda
b^\mu\partial_\mu (A_iA_j)-2\lambda^2 b^\mu b_\mu A_iA_j. \eqno(4-12)$$

       Unfortunately matters are not that simple! It is true that the
components of the latter mutiplet transform properly under Q-transformations
since each single one of the conformal Kinetic-multiplets in the definition of
(4-10;11;12) does. However, this not the case for $S$-supersymmetry since
the component, $T(\Sigma\otimes\Sigma)$, is not invariant under
$S$-supersymmetry
because the weight of $\Sigma\otimes\Sigma$  is equal to 1 instead of
${1\over2}$. Therefore, eliminating the $R$ terms is not compatible with
S-supersymmetry. Of course, recuring to a non-polynomial action [5] allows for
this possibility to occur  since each term in the infinite series expansion
compensates for the lack of $S$-supersymmetry of the  previous one as we have
already stated earlier. We are forced, then, to relinquish $S$-supersymmetry
and
implement $Q$-supersymmetry only.

     Our action is $Q$-supercovariant and is obtained by plugging-in directly
$A,\chi$ and $F$ in equation (4-9) and contracting the spacetime indices with
$\eta_{ij}$ . It has a similar form as (3-1) but it does not contain the term
linear in $S$, $S\bar{\chi}\chi$, exclusively ,which was the one which
furnished
the constraint between our physical fields in (3-1) after  elimination of
$S$. Furthermore, it contains the term ${1\over2}\bar{\chi^i}\gamma^\mu\phi_\mu
A^j$
which does not appear in (3-1). i.e; after a total derivative is performed we
end
up with ${1\over2}A\bar{\phi_\mu}\gamma^\mu\gamma^\nu\partial_\nu \chi$.
Moreover, we don't have R terms, Q-supersymmetry is linearly realized after the
elimination of F or, if we wish, after eliminating $F'$ and $S$ once we set
$F=F'+{1\over4} AS $.

    The derivatives in (4-10;11;12) contain  the spin-connection which is a
function of $e^a_\mu;\psi_\mu$ and $b_\mu$. We could have presented the
following
argument in five steps that would have allowed us us to fix the conformal-boost
invariance and set $b_\mu=0$. This occurs if, and only if, our action is
invariant
under conformal boosts.(Unfortunately it is not so; however for the sake of
completeness we shall go ahead).

     1).The Q-superconformally invariant action comprised of (4-9) after
plugging-in the values for $A;\chi$ and $F$ given by (4-10;11;12) does not
contain explicitly $ f_{a\mu}$ given by (4-5-e)( The R terms cancel out as well
as the subsequent Rarita-Schwinger terms).

     2). The fields $e^a_\mu;\psi_\mu$ and the matter multiplet, $\Sigma_C$,
are
inert under K transformations (conformal boosts).[4].

     3).Therefore, the variation of the Lagrangian with respect
K-transformations
is:

$${[{\partial L\over{\partial b_\mu}}+{\partial L\over{\partial
\omega}}{\partial \omega\over{\partial b_\mu}} +{\partial L\over{\partial
\phi_\mu}}{\partial \phi_\mu\over{\partial b_\mu}} ]\delta_K b_\mu +{\partial
L\over{\partial f_{a\mu}}}\delta_K f_{a\mu} =0.}\eqno(4-13)$$

    4). Therefore, $b_\mu$ decouples from the action if it is invariant under
conformal boosts .

    5). If $b_\mu$ decouples and , if our action was indeed invariant under
Weyl
scalings, this must be a signal that there is no need to use conformally
covariant derivatives with respect to dilatations since the action was already
Weyl invariant to begin with.
Having followed the above five-step argument we can infer that for those
actions
which are K-invariant we can fix the conformal-boost invariance and set
$b_\mu=0$.

         This is all fine but is our action ( the one given by eqs.
4-9;4-10;4-11;4-12 )
$K$-invariant? The answer is no. The $b_\mu$ terms do not decouple. We will
relegate the discussion of these terms for the Appendix. Since the presence of
these terms is harmless for the rest of the forthcoming discussion and results
we will postpone, for the time being, the discussion of the $b_\mu$ terms until
the Appendix.

         Therefore, we have constructed an action which could not have been
obtained by direct Poincare tensor calculus methods. i.e; invariant under  $Q$
but not $S$-transformations. This was the main reason why the "no-go" theorem
was
not quite correct : $S$-supersymmetry cannot be implemented in a
linearly-realized way in the absence of $R$ terms and the Poincare tensor
calculus does not even yield a linearly-realized supersymmetry for the Kinetic
terms to begin with!. An example of a multiplet which transforms properly
under the $''Q+S''$ sum rule but not separately under $Q$ nor $S$
supersymmetry is  the particular Poincare-Kinetic multiplet.[4]:

$$T_p(\Sigma_p)=(F; \not D^c\chi (\lambda ={1\over2}); \Box^c A -{3\over4}FS
).$$

This multiplet is "almost" identical to the superconformal one were it
not for the $-{3\over4}FS$ term. The last component of a Poincare multiplet
is $F'=F-{1\over2}\lambda AS$, where in the case above we have
 $\lambda=1+{1\over2}$ for $T_p(\Sigma_p).$
Therefore, one can see that it transforms properly under the "Q+S" rule but
not separately under both Q and S transformations.

        Now we turn to the supersymmetrization of the quartic terms. Why do we
need to do this if, perhaps, we could bypass it by working directly with the
action containing the cosmological constant? What happens is that we don't
retrieve the cosmological constant after eliminating $S$ from an action
comprised
of (4-9), where we use for kinetic terms solely the piece,
$\Sigma\otimes{T(\Sigma)}$ , and the one constructed from the "constant"
Poincare supermultiplet: $\Sigma_p =(1,0,0)$. i.e;  one gets negative powers of
the A field.

 We  proceed now to supersymmetrize the $L_4$ terms. One cannot
obtain a superconformally invariant action (not even Q-invariant)  now because
these terms do not have the net conformal weight of $\lambda =2$ as  the
kinetic
terms had. (We refer to the weight of the first component of a multiplet so
that  $F$  has dimension three). For this reason we have to introduce the
following coupling function, a multiplet, that has no dynamical degrees of
freedom but which serves the purpose of rendering the quartic-derivative terms
with an overall dimension three to ensure that our action is in fact
dimensionless. We refrained from doing this sort of "trick" in the case of the
kinetic terms because such terms are devoid of a dimensional coupling constant.
The Dolan and Tchrakian's action contains an arbitrary constant in front of the
quartic pieces and it is only the ratio between this constant and the
dimensionless constant in front of $L_2$ which is relevant. This constant must
have dimensions of $(length)^{+3}$ since we have an extra piece of dimension
three stemming from the term ,$(\partial_\mu  A)^2 $.

     Let us, then, introduce the coupling-function-multiplet

$$\Sigma_0=(A_0;\chi_0;F_0)$$
whose Weyl weight is equal to $-3$ so that the tensor product of $\Sigma_0$
with
the following multiplet, to be defined below, has a net conformal weight ,
$\lambda=2$ as it is required in order to have $Q$-invariant actions.

Lets introduce the following multiplet
$$K^{ijkl}_{\mu\nu\rho\tau}\eta_{ij}\eta_{kl}=K(\Sigma^i_\mu;\Sigma^j_\nu)
\otimes{T[K(\Sigma^k_\rho;\Sigma^l_\tau)}] +({ij\leftrightarrow kl})
and({\mu\nu\leftrightarrow\rho\tau})-$$

$$T[K(\Sigma^i_\mu;\Sigma^j_\nu)\otimes
{K(\Sigma^k_\rho;\Sigma^l_\tau)}]\eta_{ij}\eta_{kl}$$
where the definition of $K(\Sigma,\Sigma)$ is:

$$K(\Sigma ,\Sigma)=\Sigma^i_C \otimes T_C(\Sigma^j_C ) + T_C(\Sigma^i_C
)\otimes \Sigma^j_C -
   T_C( \Sigma^i_C \otimes \Sigma^j_C).$$
This multiplet is the adequate one to retrieve (2-7) at the bosonic level and
also the
one which ensures that the $R$ terms do cancel from the final
answer. This is indeed the case as it was shown in eqs- (4-10;11;12). After a
tedious
calculation we obtain the components of the supersymmetic-quartic-derivative
terms: (Again, notice the $b_\mu$ terms which must be present because we
have no longer conformal-boost invariance):

$${A=\bar{\chi^{ij}}\chi^{kl}.}\eqno(4-14)$$

$$\chi=-\not \partial (\bar{\chi^i}\chi^j)\chi^{kl} + ({i,j\leftrightarrow
k,l})
-{1\over2}\bar{\chi^{ij}}\chi^{kl}\gamma^\mu\psi_\mu +({i,j\leftrightarrow
k,l})$$
$${+\chi^{ij}F^{kl} +({i,j\leftrightarrow k,l}).}\eqno(4-15)$$

$$F=\bar{\psi^\mu}\partial_\mu
(\bar{\chi^i}\chi^j)\chi^{kl}+{1\over4}\bar{\psi^\mu}\psi_\mu\bar{\chi^{ij}}\chi^{kl}
$$
$$+\bar{\chi^{ij}}[-2\not D\chi^{kl}
+{1\over2}\gamma^\nu\gamma^\mu\psi_\mu\partial_\nu
(\bar{\chi^k}\chi^l)-{1\over4}\chi^{kl}\bar{\psi_\mu}\gamma^\mu\gamma^\nu\psi_\nu
+{1\over2}\gamma^\mu\psi_\mu F^{kl}]+$$
$$F^{ij}F^{kl}-2g^{\mu\nu}\partial_\mu (\bar{\chi^i}\chi^j)
\partial_\nu(\bar{\chi^k}\chi^l)+$$
$${{1\over2}\bar{\chi^{ij}}\gamma^\mu\phi_\mu\bar{\chi^k}\chi^l +
({ij\leftrightarrow kl})}$$
$$+\lambda e^{-1}\partial_\nu (eg^{\mu\nu}b_\mu
\bar\chi_i\chi_j\bar\chi_k\chi_l)+\lambda b^\mu\partial_\mu
(\bar\chi_i\chi_j\bar\chi_k\chi_l)
-2\lambda^2 b_\mu b^\mu \bar\chi_i\chi_j\bar\chi_k\chi_l .  \eqno(4-16)$$

Where we have used the abbreviations $\chi^{ij}$ and
$F^{ij}$ already given in (4-10;11;12).

      Notice the similarity between (4-14;15;16 ), above, and (4-10;11;12 ) in
form and in the values of the coefficients.This is a sign of consistency. We
need to take the tensor product of the latter multiplet given above and the
coupling-function multiplet:
$$\Sigma_0\otimes(A^{ijkl};\chi^{ijkl};F^{ijkl})=(A_0A^{ijkl};A_0\chi^{ijkl}+\chi_0A^{ijkl};
A_0F^{ijkl}+F_0A^{ijkl} -\bar\chi_0\chi^{ijkl})$$

   The complete Q-supersymmetric extension of $L_4$ requires adding terms
which result as permutations of ${ijkl\leftrightarrow ilkj\leftrightarrow
kjil\leftrightarrow klij}$ keeping $\eta_{ij}\eta_{kl}$ fixed.

We have not finish yet. One might ask the natural question : What does
the above quartic-action has to do with Dolan and Tchrakian's action? More
precisely, how do we interpret and/or dispense of the extra fields which
comprised the "coupling"function? To answer this question properly we must
first
ask ourselves what are the requirements to have a spinning membrane
action.(Q-spinning in our case). These are:

1).Linearly realized supersymmetry( Q-supersymmetry).
2). Absence of R terms.
3).Polynomial in the fields.
4).Eliminating the auxiliary fields, ${\partial L\over {\partial F^i}}=0$,
and setting the Fermi fields to zero we must recover Dolan and Tchrakian's
action. Furthermore, the order in which we perform this should yield identical
results: set Fermi fields to zero and eliminate auxiliary fields or viceversa.
This is the content of subsection {\bf 4.2} where it is shown that, in
fact, these four requirements are satisfied.

        To conclude we have $Q$-supersymmetrized the Weyl-covariantized  Dolan
and Tchrakian's action. The
kinetic terms and quartic terms are $Q$-invariant by construction. The latter
ones were Q-invariant with the aid of an extra multiplet, the
"coupling"function multiplet whose weight is precisely equal to -3 to ensure
that our action is dimensionless and scale invariant. After eliminating the
 auxiliary fields, setting the
Fermi fields to zero, and fixing the dilational gauge invariance, we
retrieve the
 Dolan and Tchrakian Lagrangian. The main
point of this paper is to show that one can only have a $Q$-spinning membrane,
solely, if we wish to satisfy all of the requirements listed before.
$"Q+S"$ invariance can only be implemented in non-polynomial actions as Rocek
and
Lindstrom showed [5].
   After subsection {\bf 4.2} we turn to the discussion concerning the presence
of the $b_\mu$ terms
which is crucial since now we do not have at our disposal the possibility of
fixing the $K$-invariance to set $b_\mu$  =0. We have decided to include this
discussion in the following Appendix because the whole essence of
section IV is based on $Q$-invariance.

\subsection{\bf Determination of $F^i,A_o,\chi_o,F_o$ and $b_\mu$.}

  After eliminating the auxiliary fields via their equations of motion and
setting the fermions to zero we must recover our initial Weyl-covariantized
Dolan-Tchrakian action (WCDT). Furthermore, the order in which perform this
must
yield identical results. Setting the fermions to zero, first, and eliminating
the $F^i$ field, per example, yields $F^i=0$ since we don't wish to generate
constraints amongst the matter fields. The $F_o A^{ijkl};\chi_o \chi^{ijkl}$
terms vanish in this limit and we are left with the bosonic pieces in
$A_oF^{ijkl}$. Similar conclusions hold for the quadratic terms as well.
Lets vary the $F^i,F_o$ fields and check that after the fermions are set to
zero we recover the WCDT action. We assume always that the target spacetime
indices are contracted with $\eta_{ij}\eta_{kl}$ and their permutations are
included. The variations with respect $F_o;F^i$ are :
$$ A^{ijkl}=\bar {\chi}^{ij}\chi^{kl}=0 \eqno (4-17)$$
$$1/2 A_o\bar\psi_\mu\gamma^\mu (\partial\chi^{ijkl}/F^i)+A_o\partial
F^{ijkl}/\partial F^i -$$
$$\bar {\chi}_o \partial \chi^{ijkl}/\partial F^i +\delta L_2 /\delta F^i =0
\eqno(4-18)$$
In the last equation we used $\chi^{ij}=0$ which is a solution of (4-17)
yielding  $F^i$ in terms of the matter fields, $A^i,\chi^i$ from
eq-(4-11). After setting the fermions to zero, the $F^i=0$ and we recover the
WCDT action. Notice that plugging the value for $F^i$ obtained from (4-17) in
eq-(4-18) yields a constraint equation amongst $A_o,\chi_o$ and the matter
fields, but in no way whatsoever, we are constraining the latter fields!

   Having the WCDT action still doesn't provide us with the original DT action.
In the Appendix we show that if the equations of motion for the membrane
coordinates obtained from the WCDT action are, indeed, the Weyl-covariantized
form of the equations of motion of the DT action, then $D_\mu A_o =0$ follows
immediately. Therefore, $b_\mu $ is zero in the $A_o =constant$ gauge. Also
we show that if $b_\mu$ is varied the bosonic action is constrained to zero.
Another way to see why the $b_\mu$ field is not determined via its equation of
motion but from the equations which follow from the $A_o,\chi_o$ variation is
the following.

   Since supersymmetry rotates field equations into field equations for the
members of a given supermultiplet, the $A_o,\chi_o$ variations are encoded
already in the $F_o$ variation and, therefore, cannot and should not be
ignored. A variation with respect to $A_o$ yields, after using eq-(4-17) with
$\chi^{ij}=0$ as a solution, in the expression for $F^{ijkl}$ :

$$F^{ij}F^{kl} +b_\mu( terms)-2g^{\mu\nu}\partial_\mu
(\bar\chi^i\chi^j)\partial_\nu (\bar\chi^k\chi^l) =0. \eqno(4-19)$$

A variation with respect to $\chi_o$ yields :

$$1/2\psi_\mu\gamma^\mu A^{ijkl}-\chi^{ijkl} =0.\eqno(4-20)$$

It isn't difficult to verify that $\chi^{ij}=0$ is a solution of (4-20) in
consistency with (4-17) by a simple inspection of eq-(4-15) and after using
eq-(4-17).

  We can now see how the $b_\mu$ field is determined by eq-(4-19). Evenfurther,
we can also see, once more, why we couldn't supersymmetrize, directly, the DT
action. If we fix the dilational gauge invariance, $A_o=constant$, while adding
compensating gauge transformations to the $Q$ supersymmetry transformation
laws,
and if we use the embedding condition :$D_\mu A_o=0$ in equations
(4-17;4-18;4-19;4-20) we find that equation (4-19) is going to constrain,
again,
the matter fields in the $b_\mu=0$ gauge, after one substitutes the value for
$F^i$ obtained from eq-(4-17). Therefore, we must have dilational gauge
invariance in our spinning membrane !. We can still plug-in the expression
$b_\mu \sim \partial_\mu ln~A_o$ into equations (4-18;4-19) yielding
$A_o,\chi_o$ in terms of the matter fields.

  We have seen how the fields $F^i;A_o,\chi_o;b_\mu$ are tightly constrained
through the use of the above equations. The $F_o$ field is undetermined
however, after the fermions are set to zero, the $F_o A^{ijkl}$ vanishes
in any case.

 Therefore, the elimination of the auxiliary fields constrains the full quartic
supersymmetric piece to vanish on shell without constraining, in any way
whatsoever, the physical fields as we have shown. Despite having only a
quadratic piece in our spinning membrane we cannot forget the presence of the
quartic terms which appears when the $F^i$ field is solved via equation (4-17)
and when the $b_\mu$ field must obey eq-(4-19) as well. The fact that $L_4$
vanishes on shell might be topological in origin. These zero actions are
important in Topological Quantum Field theories. This should be investigated
further. Of course, we cannot say that the quartic terms of the WCDT action,
after setting the fermions to zero in eq-(4-19), have to vanish. This is
because
eq-(4-19) was due to supersymmetry which is broken after the fermions are set
to zero! We must remember that $A_o;\chi_o$ were varied because they were part
of the supermultiplet which contained $F_o$ and supersymmetry forced their
variation. i.e. one cannot, simultaneously, set the fermions to zero in
equation (4-19) and still use equation (4-19) because such equation ceases to
be valid as soon as supersymmetry is broken (as soon as we set the fermions to
zero).

      Another way of rephrasing this is by saying that the $A_o$ field was
used, in the first place, to Weyl covariantize the quartic terms of the DT
action and, for this reason, $A_o$ is eliminated by fixing the dilational gauge
invariance ( $A_o$ is gauged to a constant) in the WCDT action and not via its
variation. Clearly, varying $A_o$ in the WCDT action constrains the quartic
terms to vanish and one has no longer a Weyl covariant extension of the DT
action which was, in the first place, the reason why we introduced $A_o$!!!
Similar conclusions hold if we add kinetic pieces $A_1(D_\mu A_0)^2$ to the
WCDT action and one eliminates both the $A_0$ and $A_1$ fields; one will
constrain the quartic derivative terms to vanish.

   To summarize what we just said in the prevoius paragraphs : (i). After
eliminating the $F^i, F_o $ auxiliary fields from the action and
substituting their values in the $Q$ supersymmetry transformations laws
yields a $Q$ invariant action, iff, equations (4-19;4-20) are satisfied.
(ii). The field $A_o$ can be gauged to a constant but one cannot ignore the
constraint equation which arises from its variation in the full $Q$
supersymmetric action. Such constraint owes its existence to supersymmetry
and, therefore, one cannot naively set the fermions to zero and still use
equation (4-19)to falsely claim that the quartic terms of the WCDT action
are zero !.What is zero is the full equation (4-19)- which follows from
supersymmetry- and it is no longer valid as soon as the fermions are set to
zero.

    We have shown that, in fact, eliminating the auxiliary fields via their
equations of motion and setting the fermions to zero yields the WCDT action.
The spinning membrane effectively consists of a Weyl covariant quadratic piece
like it occurs in the spinning string; however the background field $b_\mu$ and
the scalar coupling, $A_o$ are tightly constrained by a set of equations which
had their origins in the quartic terms of the $Q$ supersymmetric WCDT action.
These quartic supersymmetric terms are zero on shell. We couldn't fix, first,
the dilational gauge invariance in the supersymmetric action since constrains
will reappear amongst the matter fields. For this reason one must have
dilational gauge invariance in the spinning membrane.

Having only  $Q$ supersymmetry isn't as bad as it seems. In Poincare
supegravity
one does not have $Q$ and $S$ supersymmetry invariance, separately, but it is
only a combination of $Q$ and $S$ which is preserved, the so called $Q+S$ sum
rule. Since conformal invariance was crucial in our construction it is
warranted to study the quantum case and see how conformal anomalies will yield
information about the critical dimension; presummably this might single
out eleven dimensions.

\section {APPENDIX}

         The discussion in section IV cannot be complete unless we study in
detail the behaviour of our final action due to the presence of the $b_\mu$
terms and derive from first principles the embedding condition :$D^c_\mu
A_o =0$. To begin with, we have two cases to consider:

1-. The case where $b_\mu$ decouples from the action and from the
$Q$-supersymmetry transformation laws of the action. An example of this is
the action given by eq-(4-9) for the particular case that one chooses the
multiplet
$$ \Sigma\otimes { T(\Sigma)} $$ with a net weight equal to $\lambda=2$. One
can
see, explicitly, by simple inspection of eqs-(4-1;4-2;4-3;4-4) and all of the
eqs.
  (4-5a-e)
that the $b_\mu$ terms decouple. Therefore, there is no explicit $b_\mu$
dependence in the action (4-9) and, thus, we have implemented $K$-invariance.
\bigskip
      Furthermore, we can see that under $Q$-supersymmetry , eq.(4-9)
contains the terms:
$\delta F$ yields a term $ \bar{\epsilon}\gamma^\mu (D_\mu -(2+{1\over 2})
b_\mu)\chi$ whose $b_\mu$ term  is
$$-(2+{1\over2})\bar{\epsilon}\gamma^\mu b_\mu \chi.$$
\bigskip
A factor of $-b_\mu$ cancels against the $b_\mu$ terms contained in the
$\omega(e;\psi;b_\mu)$ leaving us with a net factor of $-{3\over2}$. Whereas
the
second term of (4-9) yields , upon variation of the gravitino using (4-7), the
following term:$${1\over2}\bar{\epsilon}\gamma^\mu b_\mu\chi. $$
plus another factor of $+b_\mu$ stemming from the spin-connection leaving a
net factor of ${3\over2}$. It is clear that $b_\mu$ does also decouple from the
transformation laws. However, if these didn't , we still have $K$-invariance
which allows to set $b_\mu$=0 and everything is fine.
\bigskip
\bigskip

2-. The case when $b_\mu$ does not decouple from the action but it does from
the transformation laws to ensure $Q$-invariance ( since we can no longer
choose
the gauge $b_\mu=0$; these $b_\mu$ terms must cancel out since Q-invariance was
not broken explicitly).
   This is our case. We bring to the attention of the reader that
$ \Lambda=T(\Sigma\otimes {\Sigma})$ is not $K$-inert. The simplest way to see
this is by looking at the superconformal algebra in three dimensions.
The Jacobi identity implies :
$$ [\Lambda,[Q,K]] +[Q,[K,\Lambda]] +[K,[\Lambda,Q]] =0.$$
Since $\Lambda$ is $Q$ invariant $\Rightarrow:$
$$ [\Lambda,Q] =0 $$
Because
$$[Q,K] \sim {S}.~~[\Lambda, S]\not = 0$$ we have
$$ [K,\Lambda]\not =0.$$

   Therefore we have $b_\mu$ terms in our final expressions for the action.
The question is: How bad is this? A careful study shows that the presence of
the $b_\mu$ terms is not harmful at all. The reasoning goes as follows:
 After the elimination of the $F^i$ auxiliary fields, per example, and
setting the fermions to zero, we have terms of the form :

$$(\partial_\mu A^i-\lambda b_\mu A^i)^2 -A_o (\partial_\mu A^i-\lambda
b_\mu A^i)^4.$$

If one attempts to eliminate the $b_\mu$ through its variation
one would arrive at a zero action. Eliminating $b_\mu$ from the above
equation, after factoring $A^i$, yields an expression of the form :
$$(D_\mu A^i)-2A_o(D_\mu A^i)^3.....=0$$

A particular non trivial solution is
$$D_\mu A^i =0$$
for all values of $i=1,2,3........d$. This implies that :
$$b_\mu \sim \partial_\mu~lnA^i$$
for all values of $i=1,2......d$. Therefore the $A^i$ are constrained  to
satisfy the condition: $A^j=k_jA^1$ for $j=2,3....d$ constants $k_j$.
Since we still have at our disposal the freedom to fix the dilational gauge
invariance, we can set $A^1=constant$ forcing all of the rest $A^j$ to be
constant and, hence, the action would be constrained to vanish. This is
unaccetable. Analogous results would have been obtained upon the
elimination of $b_\mu$ from the Weyl covariant form of the DNG action as
well.

        Therefore, in view of the above arguments, the correct procedure to
follow is to
fix the local scale invariance by setting:
$$ -A_0(x)=1.$$
and, simultaneously, set  $b_\mu$  to zero. i.e; one chooses to have trivial
background-gauge field configurations ( pure gauge ones) without dynamical
degrees of freedom. This presupposes the fact that one can find a gauge where
(simultaneously) the conformal compensator can be gauged to a constant and the
$ b_\mu$ field to zero. This can in fact be achieved by choosing for gauge
parameter the quantity:
$$ \Lambda ={1\over3}ln (- A_0 )$$
It is straightforward to verify that $b_\mu$ can be gauged to zero and $ A_0$
to
-1  simultaneously. Both of these conditions can be condensed into a
single equation :
$$ D_\mu^{Weyl}(- A_0 ) = \partial_\mu ( -A_0) +3b_\mu (- A_0) =0$$
since the weight of $A_0$  is -3.

      As we promised earlier we are going to derive the constraint on $A_o$
from first principles, from  an action. Our guiding principle will be the
on-shell dilational gauge
invariance of the WCDT equations of motion : the former must be the Weyl
covariant extension of the DT equations of motion. In particular, the
contribution to the equations of motion stemming from the quartic
derivative terms of the WCDT action are of the form :

$$D_\mu (\delta S_4/\delta (D_\mu A))\sim (D_\mu A_o)(D_\mu A)^3 +A_o D_\mu
(D_\mu A)^3.$$

  The above expression is Weyl covariant as it should be. We have assumed
that there are no boundary terms in our action and that the fields vanish
fast enough at infinity, etc.... As it is usual in these variational
problems we have integrated by parts and generalized Stokes law to the Weyl
covariant case.

  Similarly, the coresponding terms associated with the DT action are :
$$\partial_\mu (\delta S_4/\delta (\partial_\mu A))\sim
\partial_\mu (\partial_\mu A)^3.$$

   If one set of equations are, indeed, the Weyl covariant extension of the
other set, then $D_\mu A_o=0$ follows immediately. Therefore, the DT action
admits a $Q$ supersymmetric extension, iff, the fields, $A_o,b_\mu$ satisfy
the embedding condition : $D_\mu A_o=0$. This completes the results of
subsection {\bf 4.2}.

To finalize this Appendix we point out that the only obstruction in setting
$b_\mu$ to zero must be topological in origin. We saw in section  III
that it was the elimination of $S$ which originated the constraint
$\bar{\chi}\chi$ =0. Such $S$ term had the same form as a fermion-condensate.
Whereas here, upon the "trade-off" $ b_\mu \rightarrow {{1\over4}\gamma_\mu S}
$, we may  encounter topological obstructions in setting $b_\mu$ =0 gobally
and, henceforth, in $Q$-supersymmetrizing the Dolan-Tchrakian action. i.e; to
obtain the exact bosonic limit from the $Q$-supersymmetric action.

It is warranted to study the topological behaviour of these 3-dim gauge fields
and see what connections these may have with  Topological Massive Gravity
\cite{deser 7},
Chern-Simmons 3-dim Gravity and with other non-perturbative phenomena in three
dimensions.

\section{ACKNOWLEGEMENTS}

We thank Yuval Ne'eman and Luis J. Boya for helpful discussions throughout
the course of this work.

\end{document}